\documentclass{article}

\usepackage{graphicx,xcolor,balance,amsthm,epsfig}          

\usepackage{amssymb,amsmath,latexsym}
\input epsf.tex

\def\und(#1){\underline{\underline{#1}}}
\def\ove(#1){\overline{\overline{#1}}}

\def\be{\begin{equation}}
\def\ee{\end{equation}}
\def\bali{\begin{align}}
\def\eali{\end{align}}
\def\bea{\begin{eqnarray}}
\def\eea{\end{eqnarray}}
\def\beas{\begin{eqnarray*}}
\def\eeas{\end{eqnarray*}}
\def\qed{{\diamondsuit}}

\newtheorem{claimo}{Claim}

\newtheorem{theo}{Theorem}

\newtheorem{rmk}{Remark}
\newtheorem{lemma}{Lemma}

\begin{document}

\title{Boundary stabilization of systems of high order PDEs arising from flexible robotics}

\author{A. Cristofaro\thanks{{Department of Computer, Control and Management Engineering, Sapienza University of Rome,  00185 Rome, Italy. \newline Email: cristofaro@diag.uniroma1.it}}, F. Ferrante\thanks{Univ. Grenoble Alpes, CNRS, Grenoble INP, GIPSA-lab, 38000 Grenoble, France. Email: francesco.ferrante@gipsa-lab.fr.}}
\maketitle
          
\begin{abstract}
The problem of stabilization of a system of coupled PDEs of the forth-order by means of boundary control is investigated. The considered setup arises from the classical Euler-Bernoulli beam model, and constitutes a generalization of flexible mechanical systems. A linear feedback controller is proposed, and using an abstract formulation based on operator semigroup theory, we are able to prove the well-posedness and the stability of the closed-loop system. The performances of the proposed controller are illustrated by means of numerical simulations.
\end{abstract}


\section{Introduction}

Mechanical systems with flexible components have recently become an important and fertile research area, due to their versatility, high speed response and low energy consumption. Several examples of such systems may be found in soft robotics, i.e. flexible manipulators \cite{liu2018distributed} or UAVs with flexible and articulated wings \cite{paranjape2013pde}. The dynamics of this class of systems are typically governed by a combination of high-order partial differential equations (PDEs), ordinary differential equations (ODEs) and a set of static boundary conditions. Coupled systems of first and second order PDEs with ODEs have been largely investigated in the literature, and tackled with different approaches, such as backstepping \cite{coron2013local, hasan2016boundary}, Lyapunov methods \cite{trinh2017design, barreau2018lyapunov} and matrix inequalities \cite{castillo2013boundary, ferrante2019boundary}. However, the research on high order PDEs systems is more fragmentary. 

The work on control and stabilization of flexible mechanical systems was initiated by \cite{chen1979energy} using a simple model based on the wave equation, and then generalized to Timoshenko beam  \cite{kim1987boundary} and Euler-Bernoulli beam models \cite{morgul1992dynamic}.
More specifically, boundary stabilization with a single input was proved by \cite{chen1987modeling} and then extended in \cite{conrad1998stabilization}, the simultaneous stabilization of orientation and deflection was investigated in \cite{morgul1991orientation} and the tracking problem was considered in \cite{nguyen2003tracking}. Observer design was also considered by several authors, see for instance \cite{wang1991observer}, \cite{nguyen2008infinite}, \cite{jiang2017robust} and the references therein.

The present paper takes inspiration from \cite{conrad1998stabilization}, and extends the exponential stabilization results to the case of a system of coupled PDEs.  While in the scalar case the proposed controller guarantees stability for any possible choice of feedback gains, in the multi-dimensional setting some non trivial problems arise and, accordingly, some matrix inequalities have to be met by the controller parameters in order to attain stabilization. Related works are \cite{mercier2014exponential}, where serially connected beams are analyzed, and \cite{barreau2018lyapunov}, where string equations are considered.

The paper is structured as follows. Section \ref{sec:model} provides the considered model and basic setup for the stabilization problem. The well-posedness and the stability of the closed-loop system is addressed in Section \ref{sec:main}, by means of semigroup theory and matrix inequalities. Numerical simulations are given in Section \ref{sec:simulazioni} to support and illustrate the theoretical findings. Concluding remarks are finally stated in Section \ref{sec:conclusioni}.

\section{Model and setup}\label{sec:model} 
We consider the PDE system
\begin{equation}\label{eq:PDE}
u_{tt}=-\Lambda u_{xxxx}\quad x\in(0,1)
\end{equation}
with boundary conditions
\begin{eqnarray}{}
\label{eq:boundary1}Mu_{tt}(1,t)&=&Fu_{xxx}(1,t)+w(t)\smallskip\\
\label{eq:boundary2}u(0,t)&=&u_x(0,t)=u_{xx}(1,t)=0
\end{eqnarray}
where $u(\cdot,\cdot):[0,1]\times[0,\infty)\rightarrow\mathbb{R}^n,\, n\geq1$, and $\Lambda, M, F\in\mathbb{R}^{n\times n}$ are system matrices. In particular, $\Lambda=\Lambda^T\succ0$ is assumed to be diagonal and positive definite, $M=M^T\succ0$ is symmetric and positive definite, while $F$ is only required to be invertible. The function $w(\cdot)\in\mathbb{R}^n$ represents a boundary control input.\smallskip\\
The uncontrolled system is not stable, and therefore the stabilization problem is worth investigation.
Following the approach of \cite{conrad1998stabilization} for the scalar case, the idea is to design the control input as a linear state feedback consisting of a first order and a forth order term. To this end,  let us consider two matrix gains $K, B$, with $\det(B)\neq0$, and define the state feedback
\begin{equation}\label{eq:controllo}
w(t)=-Ku_t(1,t)+B^{-1}u_{xxxt}(1,t)
\end{equation}
Applying this controller, and introducing the auxiliary function $$
\eta(t)=-Fu_{xxx}(1,t)+BMu_{t}(1,t),
$$
the second-order boundary condition (\ref{eq:boundary1}) becomes
$$
B^{-1}\eta_t(t)+\eta(t)+(K-BM)u_t(1,t)=0.
$$
Furthermore, let us introduce the following functional spaces\footnote{We recall that $L^2[a,b]$ stands for the space of square integrable functions over the interval $[a,b]$, and $H^{\ell}[a,b]$ is the set of functions whose derivatives, up to the $\ell$-th order, belong to $L^2[a,b]$}:
$$
\begin{array}{rl}
\mathcal{V}:=&\Big\{u:[0,1]\rightarrow \mathbb{R}^n: u\in H^2((0,1);\mathbb{R}^n),\smallskip \\
&\quad u(0)=u_x(0)=0\Big\}
\end{array}
$$
$$
\mathcal{H}:=\Big\{(u,v,\eta):u\in\mathcal{V}, v\in L^2((0,1);\mathbb{R}^n),\eta\in\mathbb{R}^n\Big\},
$$
and consider the unbounded operator $\mathcal{A}:D(\mathcal{A})\subset \mathcal{H}\rightarrow\mathcal{H}$ defined by
\begin{equation}\label{eq:operator}
\mathcal{A}\left[
\begin{array}{c}
u\smallskip\\
v\smallskip\\
\eta
\end{array}
\right]=\left[
\begin{array}{c}
v\smallskip\\
-\Lambda^{-1}u_{xxxx}\smallskip\\
-B\eta-B(K-BM)v(1)
\end{array}
\right]
\end{equation}
whose domain is given by
$$
\begin{array}{rl}
D(\mathcal{A}):=&\Big\{(u,v,\eta): u\in H^4((0,1);\mathbb{R}^n)\cap\mathcal{V}, v\in\mathcal{V}, \eta\in\mathbb{R}^n,\smallskip\\
&u_{xx}(1)=0,\quad \eta=-Fu_{xxx}(1)+BMv(1)\Big\}.
\end{array}
$$
Accordingly, the original system can be rewritten in compact form as the linear system
\begin{equation}\label{eq:sistema_astratto}
y_t=\mathcal{A}y,\quad y(0)\in\mathcal{H}
\end{equation}
with $y=(u,u_t,\eta)$.\smallskip
\begin{rmk}[\it Mechanical interpretation]
In the scalar case $n=1$, the PDE model (\ref{eq:PDE})-(\ref{eq:boundary2}) represents the state of deflection of a flexible beam with a tip mass, or a flexible robotic arm with a fixed joint position. The generalization $n\geq2$ can be interpreted as a flexible structure with multiple arms connected in parallel and coupled at the boundary.
\end{rmk}
\begin{rmk} A simpler control input, consisting of a first order linear feedback only, can also be considered
\begin{equation}\label{eq:controllo-noexpo}
w(t)=-Ku_t(1,t)
\end{equation}
However it can be shown that, already in scalar case $n=1$, such controller guarantees asymptotical stability but not exponential stability. This is further illustrated in Section~\ref{sec:simulazioni}.
\end{rmk}
\section{Exponential stabilization}\label{sec:main}
Given two positive definite diagonal matrices $P=P^T,\ S=S^T\succ0$, and setting $y=(u,v,\eta),\ \tilde{y}=(\tilde{u},\tilde{v},\tilde{\eta})$, the following inner product is well defined in $\mathcal{H}$
$$
\langle y,\tilde{y}\rangle_{\mathcal{H}}=\int_0^1(u^T_{xx}P\tilde{u}_{xx}+v^TP\Lambda^{-1}\tilde{v})dx+\eta^TS\eta.
$$
Let us point out that the matrix $P\Lambda^{-1}\succ0$ is symmetric since both $P$ and $\Lambda$ are chosen to be diagonal. The following claim will be proved next.
\begin{claimo}\label{teo:semigroup} {\it The operator $\mathcal{A}$ generates a semigroup of contractions on the space $\mathcal{H}$ for some suitable selection of the feedback gain matrices $K, B^{-1}$. }
\end{claimo}
The idea of the proof is to invoke Lumer-Phillips theorem \cite{tucsnak2009observation}, and hinges on two technical results.
\begin{lemma}\label{lemma:dissipative}
{\it The operator $\mathcal{A}$ is dissipative if there exist positive definite diagonal matrices $P$ and  $S$, and control gains $K$ and $B$ such that the matrix inequality 
\begin{equation}\label{eq:HeOm}
\Omega+\Omega^T\succ0
\end{equation}
 is satisfied, where
\begin{equation}\label{eq:omega}
\Omega=\left[
\begin{array}{cc}
F^TSBF & \quad-F^TSBK\\ \\
P-M^TB^TSBF&\quad M^TB^TSBK
\end{array}
\right].
\end{equation}}
\end{lemma}
\proof Let us pick $y=(u,v,\eta)\in D(\mathcal{A})$ and consider the product
\begin{equation}\label{eq:prodotto}
\begin{array}{ll}
\langle y,\mathcal{A}y\rangle_{\mathcal{H}}&=\displaystyle\int_0^1(u_{xx}^TPv_{xx}-vPu_{xxxx})dx\\ \\
&-\eta^TSB(\eta+(K-BM)v(1))
\end{array}
\end{equation}
Integrating by parts twice the second term in the integral one gets
$$
\begin{array}{ll}
\displaystyle\int_0^1 v^TPu_{xxxx}dx&=v^T(1)Pu_{xxx}(1)-\underbrace{v^T(0)Pu_{xxx}(0)}_{=0}\\ \\
&\displaystyle-\int_0^1v^T_xPu_{xxx}dx\\ \\
&\displaystyle=v^T(1)Pu_{xxx}(1)-\underbrace{v^T_x(1)Pu_{xx}(1)}_{=0}\\ \\
&\displaystyle+\underbrace{v^T_x(0)Pu_{xx}(0)}_{=0}+\int_0^1v_{xx}^TPu_{xx}dx
\end{array}
$$
where the boundary conditions on $u$ and $v$ have been used. By the latter identity, the integral terms in (\ref{eq:prodotto}) cancel out and, considering the explicit expression of $\eta$ as in the definition of $D(\mathcal{A})$, one is left with the algebraic condition
$$
\langle y,\mathcal{A}y\rangle_{\mathcal{H}}=-[u_{xxx}^T(1)\, v^T(1)]\Omega \left[
\begin{array}{c}
u_{xxx}(1)\smallskip\\
v(1)
\end{array}
\right]
$$
where the matrix $\Omega$ is given by (\ref{eq:omega}). In conclusion we have $\langle y,\mathcal{A}y\rangle_{\mathcal{H}}\leq0$, and thus $\mathcal{A}$ is dissipative, as long as $\mathrm{He}(\Omega)=\Omega+\Omega\succ0$.  $\hfill\qed$

\begin{rmk}
It is worthwhile to observe that when the control gains $B$ and $K$ are given,  \eqref{eq:HeOm} is a linear matrix inequality (\emph{LMI}) in the variables $P$ and $S$. In this sense, Lemma~\ref{lemma:dissipative} recasts the dissipativity analysis of the operator $\mathcal{A}$ into the feasibility problem of an LMI. The main advantage is that checking the feasibility of an LMI is a numerically tractable problem that can be efficiently solved with available software \cite{Boyd}.
\end{rmk}
\begin{lemma}\label{lemma:onto}
{\it The range of the operator $$(\lambda I-\mathcal{A}):D(\mathcal{A})\rightarrow \mathcal{H}$$ is onto for some $\lambda>0.$}
\end{lemma}
\proof Fix $\lambda>0$. Let us show that, for any $z=(f,g,\vartheta)\in~\mathcal{H}$, there exists one $y=(u,v,\eta)$ with
$$
z=(\lambda I-\mathcal{A})y,
$$
that is
$$
\begin{array}{c}
\lambda u-v=f\smallskip\\
\lambda v+\Lambda u_{xxxx}=g\smallskip\\
(\lambda I+B)\eta+B(K-BM)v(1)=h
\end{array}
$$
Merging the first two equations yields
$$
\lambda^2 u+\Lambda u_{xxxx}=\lambda f+g,
$$
while the third equation can be rearranged as
$$
\begin{array}{rl}
&-\left(\lambda I+B\right)Fu_{xxx}(1)+\lambda B(K+\lambda M)u(1)\smallskip\\=&h+B(K+\lambda M)f(1)
\end{array}
$$
The surjectivity of $(\lambda I-A)$ is therefore equivalent to the existence of a solution to
\begin{equation}\label{eq:surje}
\begin{array}{c}
\lambda^2 u+\Lambda u_{xxxx}=f^*\quad x\in(0,1)\smallskip\\
-u_{xxx}(1)+Gu(1)=h^*\smallskip\\
u(0)=u_x(0)=u_{xx}(1)=0
\end{array}
\end{equation}
where
$$
\begin{array}{c}
f^*=\lambda f+g,\\ \\ h^*=F^{-1}(\lambda I+B)^{-1}(Bh+B(\lambda M+K)f(1)),\\ \\
G=\lambda F^{-1}(\lambda I+B)^{-1}(\lambda M+K).
\end{array}
$$
Let us denote by $\gamma_{\lambda,1},...,\gamma_{\lambda,n}$ the positive entries of the diagonal matrix $\lambda^2\Lambda^{-1}$ and by $b_{\lambda,1},...,b_{\lambda,n}$ the components of the vector $\Lambda^{-1}f^*$. Let us focus then on the scalar equations
$$
u_{xxxx}^{(j)}=-\gamma_{\lambda,j} u^{(j)}+b_{\lambda,j}
$$
Denoting by $\varrho_{j}=\root{4}\of{|\gamma_{\lambda,j}|}$, the general solution of this equation is given by
$$
\begin{array}{rl}
u^{(j)}(x)&=u^{(j)}_0(x)+\int_{0}^x\kappa_{\lambda,j}(x-z)b_{\lambda,j}(z)dz\\ \\ 
u^{(j)}_0(x)&=e^{\frac{\sqrt{2}\varrho_{j}}{2}x}\!\left(\!C_{j,1}\cos\!\left(\frac{\sqrt{2}\varrho_{j}}{2}x\right)\!+\!C_{j,2}\sin\!\left(\frac{\sqrt{2}\varrho_{j}}{2}x\right)\!\right)\smallskip\\ 
&+e^{-\frac{\sqrt{2}\varrho_{j}}{2}x}\!\left(\!C_{j,3}\cos\!\left(\frac{\sqrt{2}\varrho_{j}}{2}x\right)\!+\!C_{j,4}\sin\!\left(\frac{\sqrt{2}\varrho_{j}}{2}x\right)\!\right)
\end{array}
$$
with the convolution kernel $\kappa_{\lambda,j}(x)=e_{1}^\top\exp\{\Gamma_{\lambda,j} x\}e_4$ where 
$e_1=(1,0,0,0),\ e_4=(0,0,0,1)$ and $\Gamma_{\lambda,j}\in\mathbb{R}^{4\times4}$ is the matrix
$$
\Gamma_{\lambda,j}=\left[
\begin{array}{cc}
0_{3\times1} & I_{3\times 3}\\
-\gamma_{\lambda,j}&0_{1\times 3}
\end{array}
\right]
$$
Inserting the initial and final values, conditions on the constants $C_{j,\ell}$ can be found. In particular, denoting by $\mathbf{C}\in\mathbb{R}^{4n}$ the extended vector $$\mathbf{C}=[C_{1,1}\ C_{1,2}\ \cdots\ C_{n,3}\ C_{n,4}],$$
the existence of a solution to (\ref{eq:surje}) is equivalent to the existence of $\mathbf{C}$ such that
$$
\mathbf{\Xi}(\lambda)\mathbf{C}=\mathbf{\upsilon^*}
$$
where $\mathbf{\upsilon^*}$ is obtained by rearranging the terms in the right-hand side of (\ref{eq:surje}) and where $\mathbf{\Xi}(\lambda)\in\mathbb{R}^{4n\times 4n}$ is the matrix
of coefficients obtained imposing homogeneous conditions $$
u(0)=u_x(0)=u_{xx}(1)=u_{xxx}(1)-Gu(1)=0
$$ 
in the expression for $u^{(j)}_0(x), j=1,...,n$. A sufficient condition for the surjectivity is therefore
\begin{equation}\label{eq:determ}
\det\mathbf{\Xi}(\lambda)\neq 0.
\end{equation}
It can be noticed that, by standard results on ordinary differential equations, for $G\equiv0$ the above condition is satisfied for any $\lambda>0$. Since $G\rightarrow0$ as $\lambda\searrow0$, by continuity the condition still holds for $G\neq0$ and $\lambda$ sufficiently small. On the other hand, $\det\mathbf{\Xi}(\lambda)$ is an analytic function and then, due to the isolated zeros property \cite{rudin2006real}, condition $(\ref{eq:determ})$ must hold true for almost every $\lambda>0$.
In conclusion, the range of the operator $(\lambda I-A)$ is onto for each $\lambda>0$ such that (\ref{eq:determ}) is satisfied. $\hfill\qed$
\proof{\bf of Claim \ref{teo:semigroup}} Based on Lemma~\ref{lemma:dissipative} and  Lemma~\ref{lemma:onto}, the operator $\mathcal{A}$ satisfies the hypotheses of Lumer-Phillips theorem and thus defines a semigroup of contractions on~$\mathcal{H}.\hfill\qed$

Thanks to Claim \ref{teo:semigroup} the formal linear system (\ref{eq:sistema_astratto}) is well posed, and we can go ahead proving stability.
In fact the semigroup of contractions generated by $\mathcal{A}$ is characterized by an exponential decay, as shown in the next result.
\begin{theo}\label{teo:decay}
{\it Assume that the conditions in Lemma~\ref{lemma:dissipative} hold and let $\mathcal{T}(t)$ be the semigroup of contractions generated by the operator $\mathcal{A}$ on $\mathcal{H}$. Then, there exist two positive constants $\mu,\delta>0$ such that
$$
\|\mathcal{T}(t)\|_{\mathcal{L(\mathcal{H})}}\leq\mu e^{-\delta t}\quad t\geq0
$$
where $\|\cdot\|_{\mathcal{L(\mathcal{H})}}$ stands for the norm induced by the inner product defined earlier.}
\end{theo}
\proof Setting $z(t)=(u(\cdot,t), u_t(\cdot,t),\eta(t))$, we define the energy $W(t)$ as
$$
\begin{array}{ll}
W(t)&\!\!\displaystyle=\frac12\|z(t)\|_{\mathcal{H}}^2\\ \\
&\!\!\displaystyle=\frac12\int_0^1(u_t^T(x,t)P\Lambda^{-1}u_t(x,t)+u_{xx}^T(t,x)Pu_{xx}(t,x))dx\\ \\
&\!\!\displaystyle+\frac12 \eta^T(t)S\eta(t)
\end{array}
$$where $P,S$ satisfy the matrix inequality $\mathrm{He}(\Omega)\succ0$ in (\ref{eq:omega}). Let $z(0)\in D(\mathcal{A})$; then, by the semigroup property one has $z(t)=\mathcal{T}(t)z(0)\in D(\mathcal{A})$ 
for any $t\geq0$ and
$$
\dot{W}(t)=\langle z(t),\mathcal{A} z(t)\rangle_{\mathcal{H}}=-\zeta(t)^T\Omega\zeta(t)\leq0
$$
with $\zeta(t)=(u_{xxx}(1,t), u_t(1,t))$. Consider now the Lyapunov-like functional
$$
V(t)=tW(t)+\int_0^1xu_t^T(x,t)Qu_x(x,t)dx
$$
with $Q=Q^T\succ0$ to be selected later on. By Cauchy-Schwarz and Poincar\'e inequalities \cite{brezis2010functional}, the following estimates hold true
$$
(t-c_0)W(t)\leq V(t) \leq (t+c_0)W(t),\quad t\geq0
$$
for some $c_0>0$. Differentiation of $V(t)$ along the system solution yields
$$
\begin{array}{ll}
\dot{V}(t)&\displaystyle=W(t)+t\dot{W}(t)+\int_0^1x u^T_{xt}(x,t)Qu_t(x,t)dx\\ \\
&\displaystyle-\int_0^1x u^T_x(x,t)Q\Lambda u_{xxxx}(x,t)dx.
\end{array}
$$Let us treat the two integral terms separately. By a simple integration by parts, the first term reads as
$$
\begin{array}{ll}
&\displaystyle\int_0^1x u^T_{xt}(x,t)Qu_t(x,t)dx\\ \\
=&\displaystyle \frac12u_t^T(1,t)Qu_t(1,t)-\underbrace{\frac12u_t^T(0,t)Qu_t(0,t)}_{=0}\\ \\
-&\displaystyle\frac12\int_0^1u_t^T(x,t)Qu_t(x,t)dx.
\end{array}
$$
Integrating by parts repeatedly the second term, one gets
$$
\begin{array}{ll}
&\displaystyle\int_0^1x u^T_x(x,t)Q\Lambda u_{xxxx}(x,t)dx\\ \\
=&\displaystyle u_x^T(1,t)Q\Lambda u_{xxx}(1,t)-\underbrace{u_x^T(1,t)Q\Lambda u_{xxx}(1,t)}_{=0}\\ \\ 
-&\displaystyle\int_0^1(u_x^T(x,t)+xu_{xx}^T(x,t))Q\Lambda u_{xxx}(x,t)dx\\ \\
=&\displaystyle u_x^T(1,t)Q\Lambda u_{xxx}(1,t)-\underbrace{u_x^T(x,t)Q\Lambda u_{xx}^T(x,t){\Big |}_0^1}_{=0}\\ \\
+&\displaystyle\int_0^1 u_{xx}^T(x,t)Q\Lambda u_{xx}(x,t)dx\\ \\
-&\displaystyle \underbrace{\frac{xu_{xx}^T(x,t)Q\Lambda u_{xx}(x,t)}{2}{\Big |}_0^1}_{=0}+\frac12\int_0^1 u_{xx}^T(x,t)Q\Lambda u_{xx}(x,t)dx
\end{array}
$$
Recall the standard inequalities
$$
u_x^2(1,t)\leq\int_0^1 u_{xx}^2(x,t)dx
$$
as long as $u_x(0,t)=0$,
$$
\begin{array}{cc}
&u_x^T(1,t)Q\Lambda u_{xxx}(1,t)\smallskip\\
\leq &\|Q\Lambda\| \left(\epsilon \|u_x(1,t)\|^2+\frac1\epsilon\|u_{xxx}(1,t)\|^2\right) 
\end{array}$$ and
$$
\begin{array}{cc}
&\eta^T(t)S\eta(t) \smallskip \\
\leq &2\|S\|\left(\|F\|^2\|u_{xxx}(1,t)\|^2+\|BM\|^2\|u_t(1,t)|^2\right)
\end{array}
$$
where $\epsilon>0$ is arbitrary. Now select the matrix $Q\succ0$ such that
\begin{eqnarray}
\label{eq:Q1}\frac32 \mathrm{He}(Q\Lambda)-\frac12P\succ\alpha I\smallskip\\
\label{eq:Q2}Q-P\Lambda^{-1}\succ\beta I
\end{eqnarray}
for some $\alpha,\beta>0$. Let us point out that these two conditions are always simultaneously feasible.
Putting all pieces together, the following estimate on $\dot V(t)$ is found
$$
\begin{array}{ll}
\dot{V}(t)\leq&\displaystyle-(\alpha-\epsilon\|Q\Lambda\|)\int_0^1u^T_{xx}(x,t)u_{xx}(x,t)dx\\ \\
&-\displaystyle\beta \int_0^1u^T_{t}(x,t)u_{t}(x,t)dx-\zeta^T(t)(t\Omega-\Psi)\zeta(t)
\end{array}
$$
with
$$
\Psi\!=\!{\left[
\begin{array}{cc}
\!\|S\|\|F\|^2+\frac{\|Q\Lambda\|}{\epsilon}&0\\ \\
0&\|S\|\|BM\|^2+\frac{\|Q\|}2
\end{array}
\!\right]}
$$
It can be easily verified that, selecting $\epsilon<\frac{\alpha}{\|Q\Lambda\|}$, there exists $\tau>0$ such that the right-hand side is nonpositive for any $t\geq\tau$, i.e.
$$
\dot V(t)\leq 0\quad \forall t\geq\tau.
$$
In turn this implies that
$$
W(t)\leq\frac{\tau+c_0}{t-c_0}W(0)\quad t>\max\{c_0,\tau\}
$$and thus from the identity
$$
W(t)=\frac12\|z(t)\|^2_{\mathcal{H}}=\frac12\|\mathcal{T}(t)z(0)\|^2_{\mathcal{H}}
$$it follows that $\|\mathcal{T}(t)\|_{\mathcal{L}(\mathcal{H})}<1$ for $t>\max\{c_0,\tau\}$, this being equivalent to the claimed exponential decay due to the semigroup property. In conclusion, we have proved that the controller (\ref{eq:controllo}) provides exponential stabilization.$\hfill\qed$

\section{Numerical Example}\label{sec:simulazioni}
Let us consider a system in the form (\ref{eq:PDE})-(\ref{eq:boundary2}), described by the matrices:
$$
\begin{array}{c}
\Lambda=\left[
\begin{array}{cc}
15 &0\\0&10
\end{array}
\right],\quad M=\left[
\begin{array}{cc}
0.1&0\\0&0.3
\end{array}
\right]\\ \\
F=\left[
\begin{array}{cc}
2&0.4\\-0.8&1
\end{array}
\right]
\end{array}
$$
and initial condition $$
u(x,0)=\left[0.03\sin\left(\frac{\pi x}{2}\right)\ -0.02\cos\left(\frac{\pi x}{2}\right) \right]^T.
$$
In addition, we assume that the controller gains are selected as follows:
\begin{equation}\label{eq:Gains}
K=\left[
\begin{array}{cc}
20&0\\0&20
\end{array}
\right],\quad B^{-1}=\left[
\begin{array}{cc}
0.01&0\\0&0.02
\end{array}
\right].
\end{equation}
Solving (\ref{eq:HeOm}) in Matlab\textsuperscript{\tiny\textregistered} using the YALMIP package \cite{lofberg2004yalmip} combined with the solver SDPT3, \cite{tutuncu2003solving}, one gets:
$$
P=\left[\begin{array}{cc} 0.954272 & 0\\ 0 & 0.881155 \end{array}\right], S= \left[\begin{array}{cc} 0.00197144 & 0\\ 0 & 0.0271201
\end{array}\right]$$
The simulation have been performed using a finite-difference scheme, rearranging the discretized model of a beam described in \cite[Section 5.4]{liu2018distributed} for a motorized robotic arm.

\begin{figure}[h!]
\centering
\includegraphics[width=0.9\columnwidth]{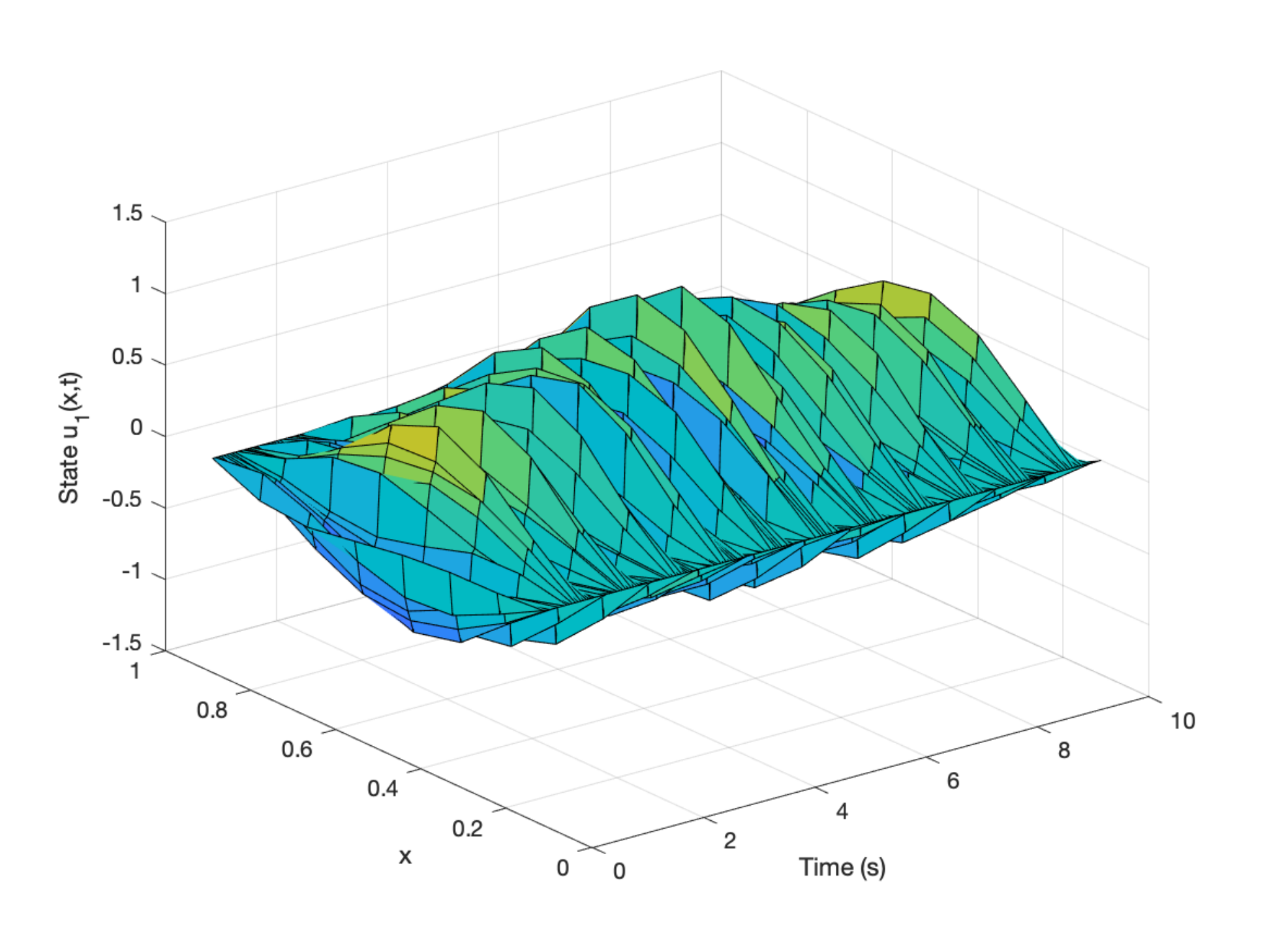}
\caption{Case a): State $u_1(x,t)$ without control input}\label{fig:fig1_1}
\end{figure}

\begin{figure}[h!]
\centering
\includegraphics[width=0.9\columnwidth]{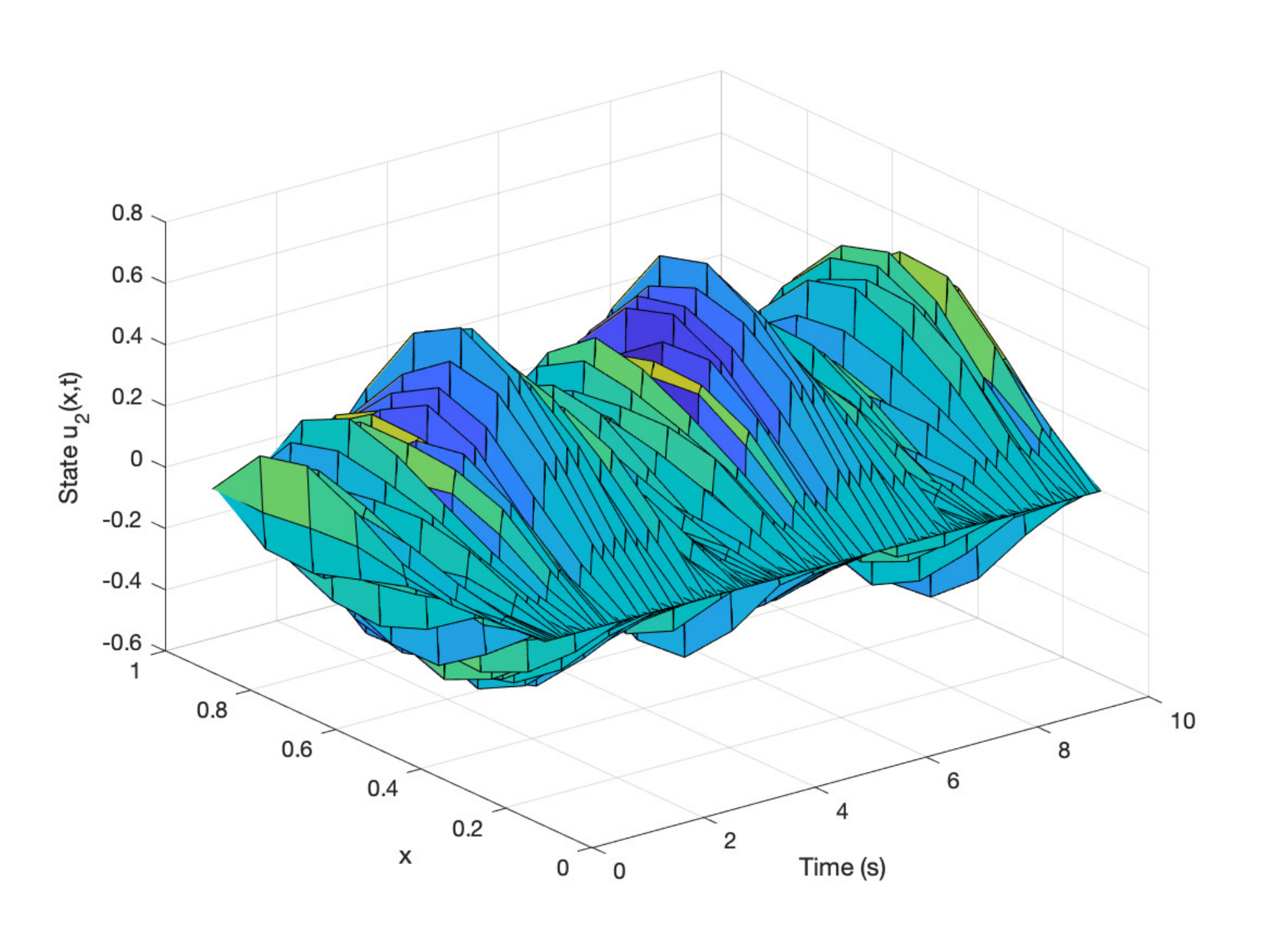}
\caption{Case a): State $u_2(x,t)$ without control input}\label{fig:fig2_1}
\end{figure}

\begin{figure}[h!]
\centering
\includegraphics[width=0.9\columnwidth]{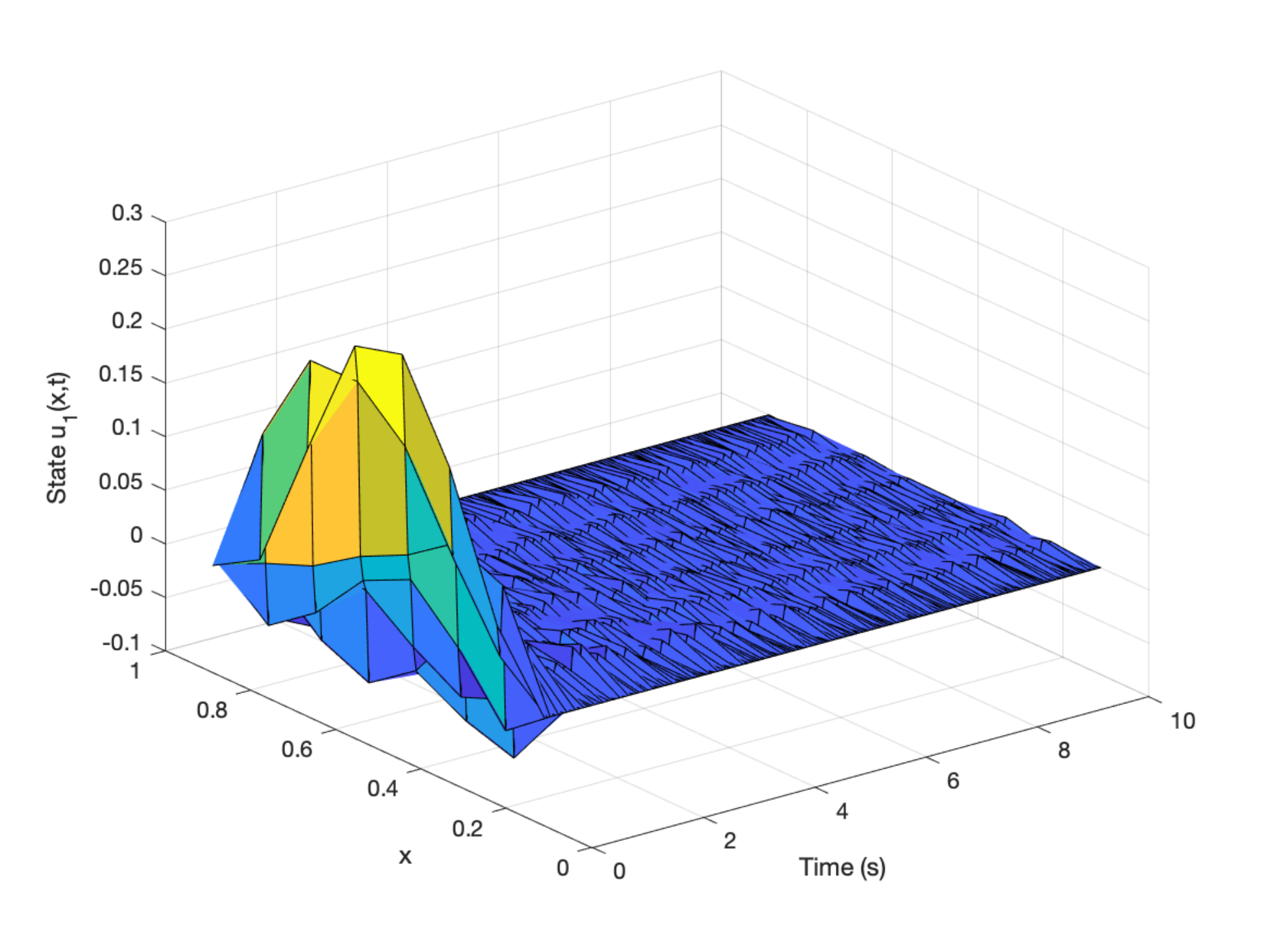}
\caption{Case b): State $u_1(x,t)$ with first order control (\ref{eq:controllo-noexpo})}\label{fig:fig1_2}
\end{figure}

\begin{figure}[h!]
\centering
\includegraphics[width=0.9\columnwidth]{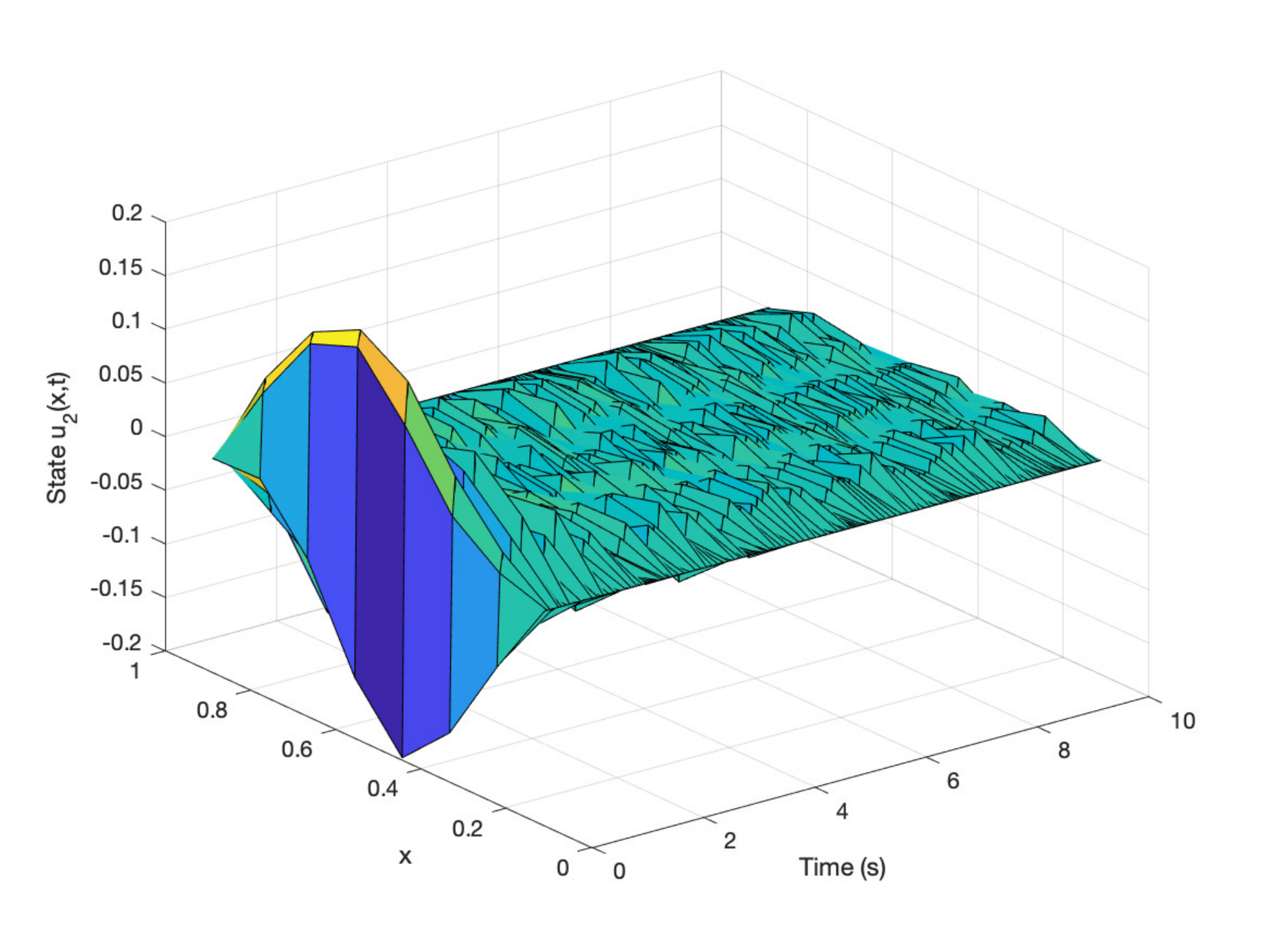}
\caption{Case b): State $u_2(x,t)$ with first order control (\ref{eq:controllo-noexpo})}\label{fig:fig2_2}
\end{figure}

\begin{figure}[h!]
\centering
\includegraphics[width=0.9\columnwidth]{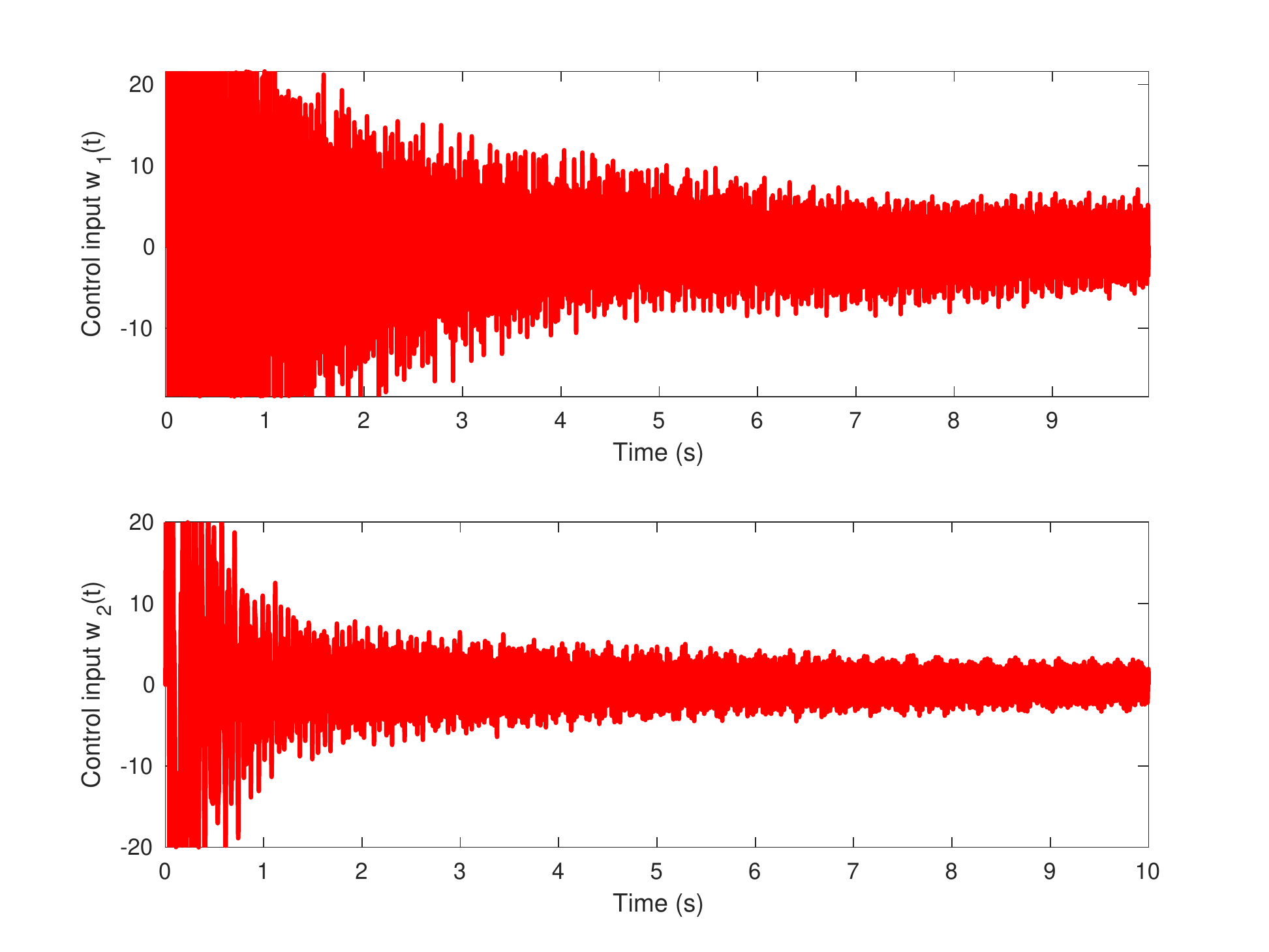}
\caption{Case b): First order control inputs (\ref{eq:controllo-noexpo})}\label{fig:fig3_2}
\end{figure}

\begin{figure}[h!]
\centering
\includegraphics[width=0.9\columnwidth]{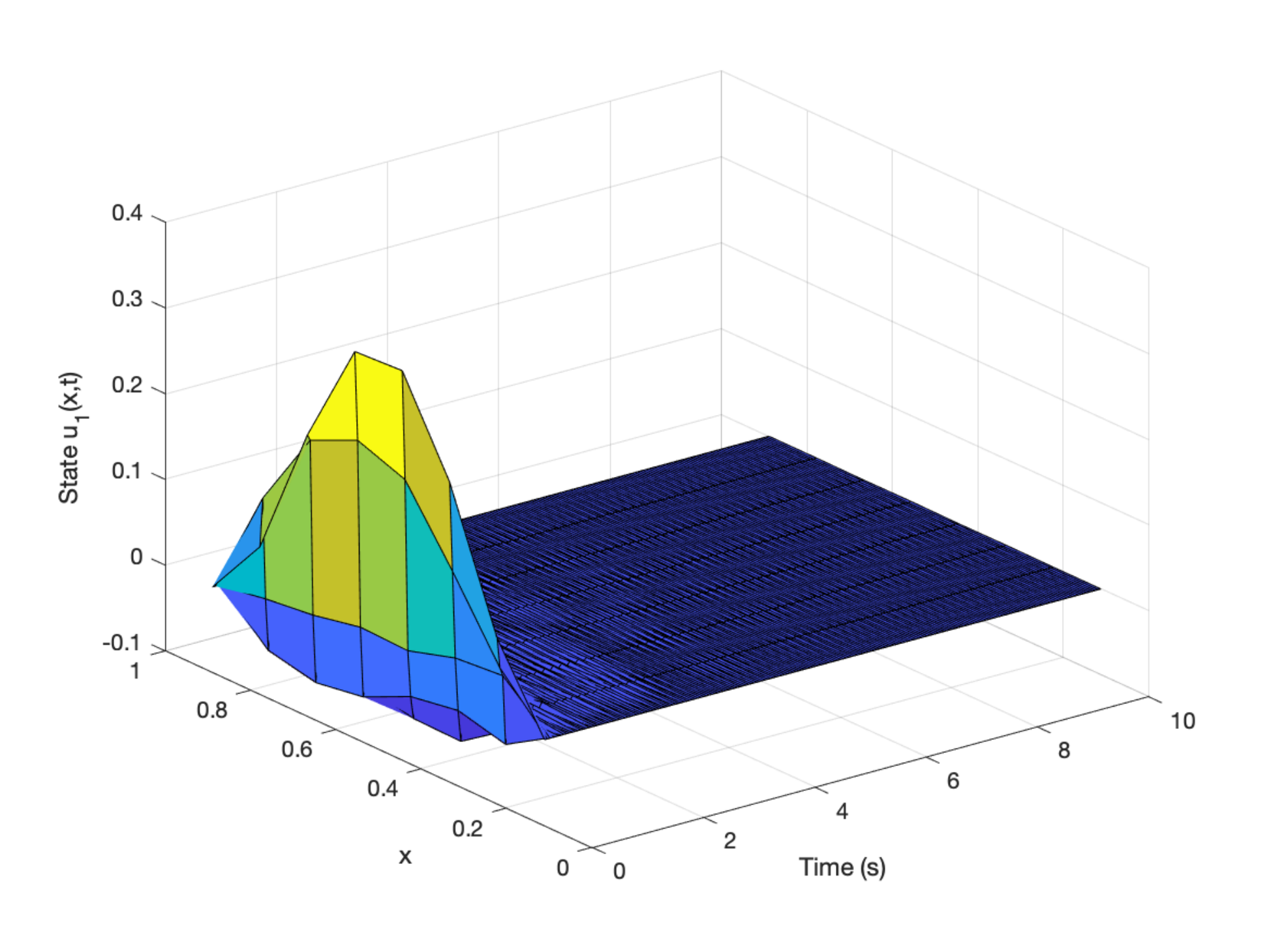}
\caption{Case c): State $u_1(x,t)$ with control input (\ref{eq:controllo})}\label{fig:fig1_3}
\end{figure}

\begin{figure}[t!]
\centering
\includegraphics[width=0.9\columnwidth]{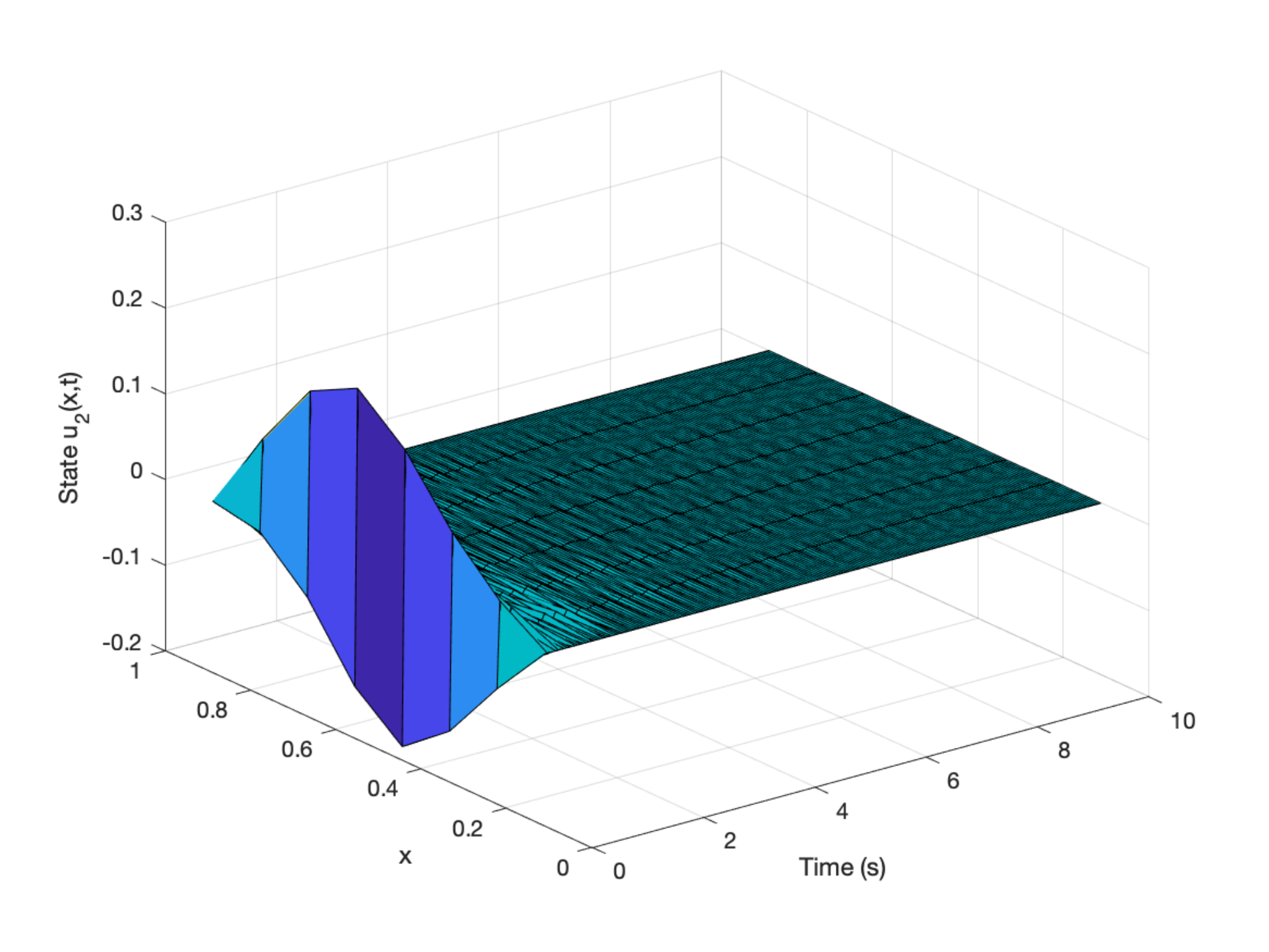}
\caption{Case c): State $u_2(x,t)$ with control input (\ref{eq:controllo})}\label{fig:fig2_3}
\end{figure}

\begin{figure}[t!]
\centering
\includegraphics[width=0.9\columnwidth]{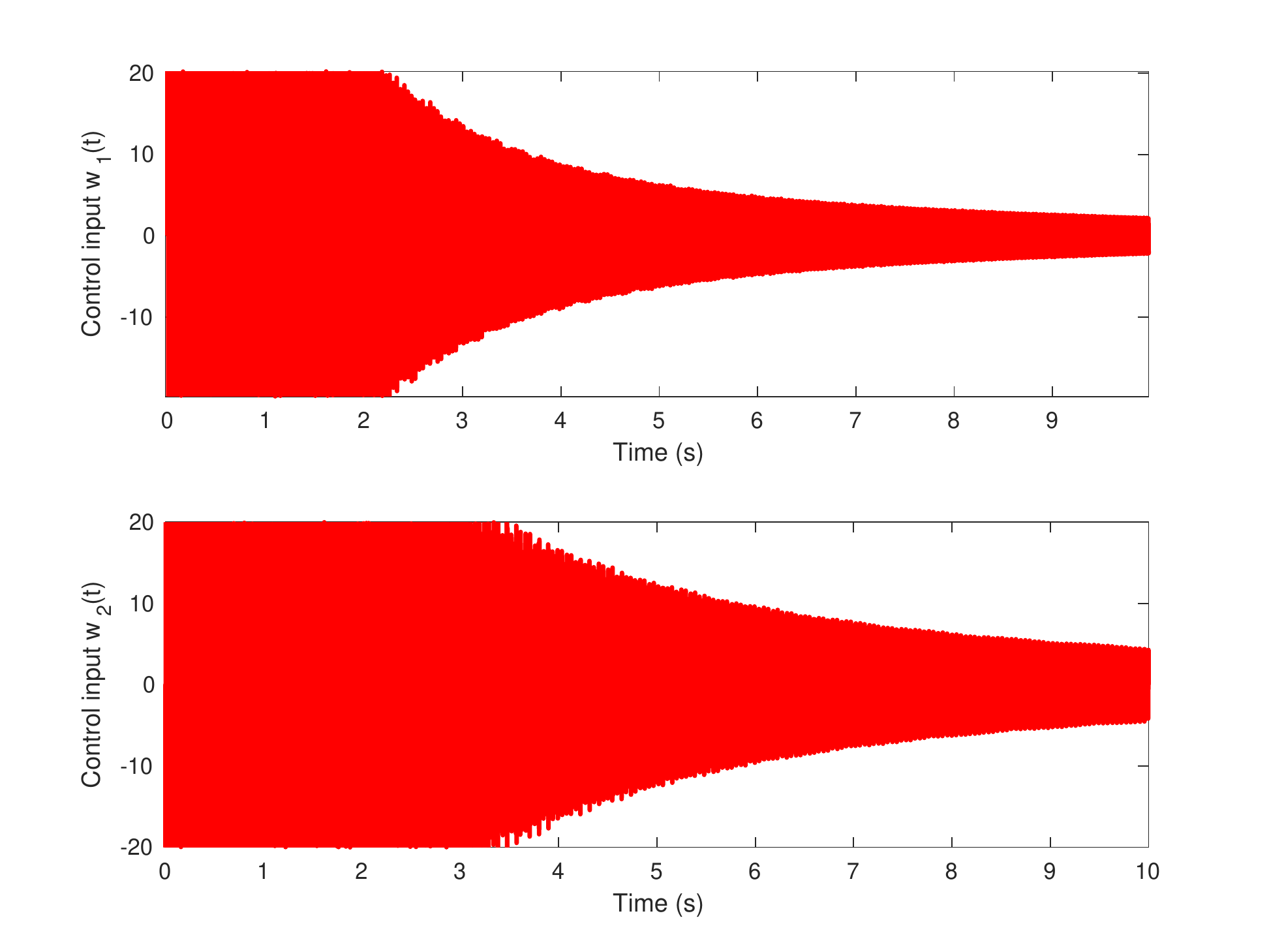}
\caption{Case c): Control inputs (\ref{eq:controllo})}\label{fig:fig3_3}
\end{figure}

Three cases have been considered: a) uncontrolled system, i.e. $w(t)\equiv0$; b) first order controller $w(t)$ given by (\ref{eq:controllo-noexpo}); c)~complete controller given by (\ref{eq:controllo}). The state evolution for case a) is depicted in Figures \ref{fig:fig1_1}-\ref{fig:fig2_1}. It is clearly visible from the plots that the initial condition is propagated and not attenuated during the evolution, this showing that the uncontrolled system is not asymptotically stable. The performances of the first order controller (\ref{eq:controllo-noexpo}) are illustrated in Figures \ref{fig:fig1_2}-\ref{fig:fig3_2}. The state of the equation slowly converges to zero, this showing that asymptotical stability is achieved. The control input is characterized by an oscillatory behaviour (see Figure \ref{fig:fig3_2}), this being induced by the initial condition and then propagated by the time derivative in the state feedback. Finally, case c) is presented in Figures \ref{fig:fig1_3}-\ref{fig:fig3_3}. The improved performances with respect to case b), i.e. exponential stability, can be appreciated in Figures \ref{fig:fig1_3}-\ref{fig:fig2_3}. On the other hand it must be noticed that, due to the presence of the higher order derivative in the controller structure (\ref{eq:controllo}), an increased level of chattering appears in the control input (see Figure \ref{fig:fig3_3}).
\smallskip\\
\smallskip\\
\smallskip\\
\smallskip\\
\smallskip\\
\smallskip\\
\smallskip\\
\section{Conclusions}\label{sec:conclusioni}
A generalization of the stabilization problem of a flexible beam with a tip mass is considered in this paper. In particular, the exponential stabilization of a system of $n\geq2$ coupled high order PDEs by means of boundary control is addressed. The problem has been tackled using operator semigroups theory, Lyapunov methods and matrix inequalities. Unlike the scalar case, some non trivial algebraic conditions arise in the case $n\geq2$, this making the stability analysis and synthesis of controllers more challenging and interesting. Sufficient conditions in the form of linear matrix inequalities are given to assess global exponentially stability of the closed-loop system.
Future research directions include the derivation of a control design algorithm based on the solution to linear matrix inequalities and the analysis of parametric uncertainties.
\bibliographystyle{plain}
\bibliography{beam}
\balance
\end{document}